\documentclass[amsmath,amssymb,twocolumn,showpacs,showkeys,superscriptaddress]{revtex4}
\usepackage{epsfig}
\usepackage{graphicx}
\usepackage{dcolumn}
\usepackage{bm}
\begin{document}
\title{Basic kinetic wealth-exchange models: common features and open problems}
\author{Marco Patriarca}
  \email{{\tt marco.patriarca [at] gmail.com}}
  \affiliation{IFISC, 
  Instituto de F\'isica Interdisciplinar y Sistemas Complejos (CSIC-UIB), 
  E-07122 Palma de Mallorca, Spain}
  \affiliation{National Institute of Chemical Physics and Biophysics, 
  R\"avala 10, 
  Tallinn 15042, Estonia}
\author{Els Heinsalu}
  \affiliation{IFISC, 
  Instituto de F\'isica Interdisciplinar y Sistemas Complejos (CSIC-UIB), 
  E-07122 Palma de Mallorca, Spain}
  \affiliation{National Institute of Chemical Physics and Biophysics, 
  R\"avala 10, 
  Tallinn 15042, Estonia}
\author{Anirban Chakraborti}
  \affiliation{Laboratoire de Math\'{e}matiques Appliqu\'{e}es aux Syst\'{e}mes,
  Ecole Centrale Paris,
  92290 Ch\^{a}tenay-Malabry, France}

\date{\today}

\begin{abstract}
We review the basic kinetic wealth-exchange models of 
Angle 
[J. Angle, Social Forces 65 (1986) 293; J. Math. Sociol. 26 (2002) 217], 
Bennati 
[E. Bennati, Rivista Internazionale di Scienze Economiche e Commerciali 35 (1988) 735], 
Chakraborti and Chakrabarti
[A. Chakraborti, B. K. Chakrabarti, Eur. Phys. J. B 17 (2000) 167], 
and of Dragulescu and Yakovenko 
[A. Dragulescu, V. M. Yakovenko, Eur. Phys. J. B 17 (2000) 723].
Analytical fitting forms for the equilibrium wealth distributions 
are proposed.
The influence of heterogeneity is investigated, 
the appearance of the fat tail in the wealth distribution
and the relaxation to equilibrium are discussed.
A unified reformulation of the models considered is suggested.
\end{abstract}
%
%
\maketitle
%

\section{Introduction}

Many scientists have underlined the importance of a quantitative approach 
in social sciences~\cite{Bouchaud2005a,Chatterjee2005b,Feigenbaum2003a,Mantegna2000a,Stauffer2005a,Stauffer2006c,Chakrabarti2006a,Taagepera2008b}.
In fact, statistical mechanics and social sciences have been always linked 
to each other in a constructive way, due to the statistical character 
of the objects of study~\cite{Ball2002a,Yakovenko2009a}.
On one hand, 
various discoveries, first made in the field of social sciences,
introduced new concepts which turned out to be relevant
for the development of statistical mechanics 
and later of the science of complex systems.
For instance, fat tails were found by Pareto 
in the distribution of wealth~\cite{Pareto1897a,Pareto1971a_reprint};
the first description of financial time series through statistical mechanics, 
made by L. Bachelier in his PhD thesis~\cite{Bachelier1900a,Bachelier1990b,Boness1967a},
also represents the first formalization of a stochastic process 
in terms of the random walk model;
large fluctuations were observed by Mandelbrot in the
time series of cotton price~\cite{Mandelbrot1963a}.
On the other hand, physics has often represented a prototype
for modelling economic systems.
For example, many works of Paul Samuelson were inspired by thermodynamics;
the analogies between physics and economics were studied 
by Jan Tinbergen in his PhD thesis entitled
``Minimum Problems in Physics and Economics''.
Recent developments of economics rely more and more 
on the theory of stochastic processes
and the science of complex systems~\cite{Arthur1999a}.

The present paper considers some models of wealth exchange 
between individuals or economical entities,
introduced independently in different fields 
such as social sciences, economics, and physics.
We refer to them as {\it ki\-ne\-tic wealth-ex\-chan\-ge mo\-dels} (KWEM), 
since they provide a description of wealth flow in terms of
stochastic wealth exchange between {\it agents},
resembling the energy transfer
between the molecules of a fluid~\cite{Chatterjee2005b,Hayes2002a,Chatterjee2007b}.
In order to maintain the discussion at a fundamental level,
we limit ourselves to the following simple KWEMs: 
those introduced by Angle (A-models)~\cite{Angle1983a,Angle1986a,Angle2002a,Angle2006a}, 
Bennati~(B-model)~\cite{Bennati1988a,Bennati1988b,Bennati1993a},
Cha\-kra\-borti and Chak\-ra\-bar\-ti~(C-\-mo\-del)~\cite{Chakraborti2000a}, 
and by Dragulescu and Ya\-ko\-ven\-ko~(D-mo\-del)~\cite{Dragulescu2000a}.
The goal of the paper is to discuss their general common features,
formulation, and stationary solutions for the wealth distribution.
We consider a heterogeneous KWEM, in order to illustrate 
how a simple KWEM can generate realistic wealth distributions.
We also clarify some relevant issues, recently discussed in the literature,
concerning the relaxation to equilibrium and the appearance of
a power law tail of the equilibrium distribution in heterogeneous models.

A noteworthy difficulty in the study of wealth or money exchanges based on
a kinetic approach had been pointed out by Mandelbrot~\cite{Mandelbrot1960a}:
\begin{quote} 
... there is a great temptation to consider the exchanges of money 
which occur in economic interaction as analogous to the exchanges of energy 
which occur in physical shocks between gas molecules... 
Unfortunately the Pareto distribution decreases much more slowly 
than any of the usual laws of physics...
\end{quote}
The problem referred to in this quotation is that 
the asymptotic shape of the energy distributions of gases 
predicted by statistical mechanics usually have the Gibbs form or 
a form with an exponential tail.
The real wealth distributions, instead, exhibit a Pareto power law 
tail~\cite{Pareto1897a,Pareto1971a_reprint,Levy1997a,Fujiwara2003a,Aoyama2003a},
\begin{eqnarray}
  f(x) \sim 1/x^{1+\alpha} \, ,
  \label{powerlaw}
\end{eqnarray}
with $1 < \alpha < 2$.
However, it has become clear that
(a) the actual shapes of wealth distribution at intermediate values of wealth 
are well fitted by a $\mathrm{\Gamma}$- or an exponential distribution~\cite{Angle1986a,Dragulescu2001a,Dragulescu2001b,Ferrero2004a},
so that they can be reproduced also by simple KWEMs 
with homogeneous agents (see Sec.~\ref{homogeneous});
(b) KWEMs with suitably \emph{diversified} agents can generate
also the power law tail of the wealth distribution~\cite{Iglesias2004a,Chatterjee2005b,Chatterjee2007b} (see Sec.~\ref{heterogeneous}).
This has opened the way to a simple, quantitative approach in modelling
real wealth distributions as arising from wealth exchanges 
among economical units.

The paper is structured as follows:
In Sec.~\ref{model} a general description of a KWEM is given.
In Sec.~\ref{homogeneous} the homogeneous A-, B-, C, 
and D-models are discussed.
Explicit analytical fitting forms for the equilibrium wealth distributions
are given.
In Sec.~\ref{heterogeneous} we discuss the influence of heterogeneity, 
taking the heterogeneous C-mo\-del as a representative example.
In this respect, we analyze the mechanism leading to a robust power law tail.
Some issues concerning the convergence time scale of the model
and the related finite cut-off of the power law are discussed. 
In Sec.~\ref{reformulation} a unified reformulation 
of the A-, C-, and D-models is suggested, 
which in turn naturally lends itself to further generalizations.
An example of generalized model is worked out in detail.
Conclusions are drawn in Sec.~\ref{conclusion}.

\section{General structure}
\label{model}

In the models under consideration the system is assumed to be made up
of $N$ agents with wealths $\{x_i \ge 0\}$ ($i = 1, 2,\dots, N$).
At every iteration an agent $j$ exchanges a quantity $\Delta x$ 
with another agent $k$ chosen randomly.
The total wealth $X=\sum_{i} x_i$ is constant as well as 
the average wealth $\langle x \rangle = X/N$.
After the exchange the new values $x_j'$ and $x_k'$ are ($x_j', x_k' \ge 0$)
\begin{eqnarray}
  x_j' &=& x_j - \Delta x \, ,
  \nonumber \\
  x_k' &=& x_k + \Delta x \, .
  \label{basic0}
\end{eqnarray}
Here, without loss of generality, the minus (plus) sign has been chosen 
in the equation for the agent $j$ ($k$). 
The form of the function $\Delta x = \Delta x(x_j, x_k)$ defines
the underlying dynamics of the model.

In KWEMs, agents can be characterized by 
an exchange parameter $\omega \in (0, 1]$
which defines the maximum fraction of the wealth $x$ 
that enters the exchange process.
Equivalently, one can introduce the saving parameter $\lambda = 1 - \omega$, 
with value in the interval $[0, 1)$, representing the minimum fraction 
of $x$ preserved during the exchange.
The parameter $\omega$ ($\lambda$) also determines the time scale 
of the relaxation process as well as 
the mean value $\langle x \rangle$ at equilibrium~\cite{Patriarca2007a}.
If the value of $\omega $ ($\lambda$) is the same for all the agents, 
the model is referred to as {\it homogeneous} (see Sec.~\ref{homogeneous}).
If the agents assume different values $\omega _i$ ($\lambda _i$) 
then the model is called {\it heterogeneous} (see Sec.~\ref{heterogeneous}).
Homogeneous models can reproduce the shape of the $\mathrm{\Gamma}$-distribution 
observed  in real data at small and intermediate values of the wealth.
For $\omega < 1$ ($\lambda > 0$), they have the self-organizing property 
to converge toward a stable state with a wealth distribution 
which has a non-zero median, differently from a purely exponential distribution.
Models with suitably diversified agents can reproduce also 
the power law tail (\ref{powerlaw}) found in real wealth distributions.

In actual economic systems the total wealth is not conserved 
and a more faithful description should be used.
It is therefore interesting to observe how 
the closed economy models considered here, 
in which $\sum_i x_i$ is constant, 
provide realistic shapes of wealth distributions.
This suggests that the main factor determining the wealth distribution
is the \emph{wealth exchange}.

When the variation of wealths is not due to an actual exchange 
between the two agents but the quantity $\Delta x$ 
is entirely lost by one agent and gained by the other one, 
the model is called \emph{unidirectional}.
Furthermore, it is possible to conceive multi-agent interaction models,
not considered here, in which a number $M > 2$ of agents enter each trade.
Then the evolution law has the more general form $x_i' = x_i + \Delta x_i$, 
with $i=1,\dots,M$, $\sum_{i=1}^M \Delta x_i = 0$, 
and the $\Delta x_i$ depending somehow on the wealths $x_i$
of the $M$ interacting agents.

\section{Homogeneous models}
\label{homogeneous}

\subsection{A1-model} \label{angle1}

Here we consider the model introduced by John Angle in 1983 
in Refs.~\cite{Angle1983a,Angle1986a},
referred to as A1-model
(a different model of Angle, the One-Parameter Inequality Process,
referred to as A2-model, is consider in Sec.~\ref{angle2} below).
The A-models are inspired by the surplus theory of social stratification and
describe how a non-uniform wealth distribution arises from wealth exchanges 
between individuals.

The A1-model is unidirectional and its dynamics is highly nonlinear.
The dynamical evolution is determined by Eqs.~(\ref{basic0}) with $\Delta x$ given as
\begin{eqnarray}
  \Delta x =
  \epsilon \, \omega \, [ \eta_{j,k}  \, x_j - (1-\eta_{j,k}) \, x_k ]
  \, .
  \label{basic_angle}
\end{eqnarray}
Here $\epsilon$ and $\eta_{j,k}$ are random variables. 
The first one is a random number in the interval $(0,1)$, which 
can be distributed  either uniformly or 
with a certain probability distribution $g(\epsilon)$, 
as in some generalizations of the basic A1-model~\cite{Angle1983a}.
The second one is a random dichotomous variable responsible 
for the unidirectionality of the wealth flow as well as 
for the nonlinear character of the dynamics. 
It is a function of the difference between the wealths of the interacting agents
$j$ and $k$, $\eta_{j,k} \equiv \phi(x_k - x_j)$,
assuming the value $\eta_{j,k} = 1$ 
with probability $p_0$ for $x_j > x_k$ 
or the value $\eta_{j,k} = 0$ with probability $1-p_0$ for $x_k > x_j$.
The value $\eta_{j,k}=1$ produces a wealth transfer 
$|\Delta x| = \epsilon \, \omega \, x_j$ 
from agent $j$ to $k$, while the value $\eta_{j,k}=0$ corresponds 
to a wealth transfer $|\Delta x| = \epsilon \, \omega \, x_k$ from $k$ to $j$.

\begin{figure}
\resizebox{0.90\columnwidth}{!}{
\includegraphics{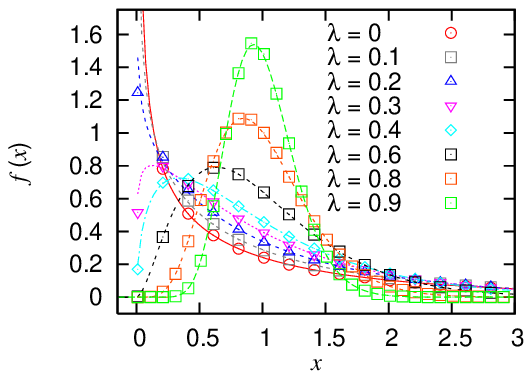}}\\
\resizebox{0.9\columnwidth}{!}{
\includegraphics{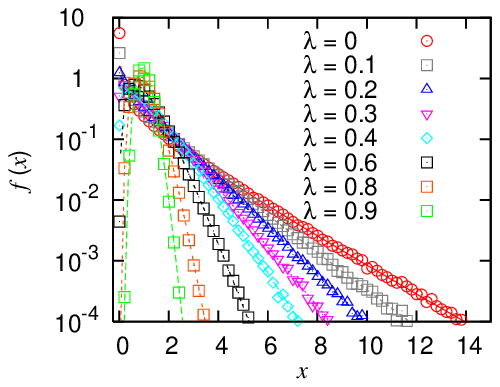}
}
\caption{
Equilibrium wealth distribution for the basic version of the A1-model 
defined by Eqs.~(\ref{basic_angle}), for the  case $p_0=1/2$: 
results of numerical simulations (symbols) 
and fitting functions Eqs.~(\ref{JA1}) (curves) 
for different values of the saving parameter $\lambda = 1 - \omega$ 
in linear (above) and semi-log (below) scale.
The value of $n$ is governed by Eq.~(\ref{JA2}).
In this simulation the average wealth is $\langle x \rangle=1$.
}
\label{fig1}
\end{figure}

A special case of the A1-model is that with symmetrical interaction,
obtained for $p_0 = 1/2$. 
Notice that for this value of $p_0$ 
the random variable $\eta_{j,k} \equiv \eta$ 
becomes independent of $x_j$ and $x_k$.
We have studied this particular case through numerical simulations 
for various values of the saving parameter $\lambda$.
The system considered was made up of $N=10^5$ agents
with equal initial wealths $x_{0i}^{\,} = 1$
and the transactions were performed until equilibrium was reached.
The equilibrium distributions in Fig.~\ref{fig1} were obtained 
by averaging over $10^5$ different runs. 
They are well fitted by the $\mathrm{\Gamma}$-distribution
\begin{eqnarray}
  \label{JA1}
  f(x) &=& \beta\,\gamma_n(\beta x)
  = \frac{\beta}{\Gamma(n)} \, (\beta x)^{n-1} \exp( - \beta x ) \, ,
\end{eqnarray}
where
\begin{eqnarray}
  \label{JA2}
  \beta^{-1} 
  &=& \langle x \rangle / n \, ,
  \\
  n 
  &\equiv& \frac{D}{2} 
  = \frac{1 + 2\lambda}{2(1 - \lambda)}
  =   \frac{3}{2 \omega} - 1 \, .
\end{eqnarray}
Since $\lambda = 1 - \omega \in [0,1)$, the parameter $n$ 
is a real number in the interval $[1/2, \infty)$.
Notice that from Eqs.~(\ref{JA1}) and (\ref{JA2}) 
it follows that for $n<1$, i.e., for $\lambda < 1/4$ 
($\omega > 3/4$), the $\mathrm{\Gamma}$-distribution diverges for $x \to 0$, 
as visible in Fig.~\ref{fig1} for the cases 
$\lambda \!=\! 0$, $\lambda \!=\! 0.1$, and $\lambda \!=\! 0.2$.
For the critical value $\lambda = 1/4$ ($\omega = 3/4$),
separating the distributions which diverge from those which go to zero
for $x \to 0$, an exponential distribution is obtained,
\begin{eqnarray}
  \label{exp}
  f(x) = \beta\,\gamma_1(\beta x) = \beta \exp(-\beta x) \, .
\end{eqnarray}

The A1-model has a simple mechanical analogue if the quantity $D = 2n$ 
defined in Eq.~(\ref{JA2}) is interpreted 
as an effective dimension for the system
and $\beta^{-1}$ as a temperature. 
It is easy to check that the distribution 
$\gamma_n(\beta x) \equiv \gamma_{D/2}(\beta x)$ 
given by Eq.~(\ref{JA1}) is the equilibrium distribution 
for the kinetic energy of a perfect gas in $D$ dimensions 
as well as for the potential energy of a $D$-dimensional harmonic oscillator 
or a general harmonic system with $D$ degrees of freedom.
This definition of effective dimension is consistent 
with the equipartition theorem, since
\begin{eqnarray}
  \langle x \rangle = n\,\beta^{-1} = D\beta^{-1}/2 \, ,
\end{eqnarray}
see Ref.~\cite{Patriarca2004a} for details.

\subsection{A2-model} \label{angle2}

The One-Parameter Inequality Process model, here referred to as A2, 
is another model introduced by John Angle and is described in detail in 
Refs.~\cite{Angle2002a,Angle2006a}.
It differs from the A1-model considered above in that it only employs 
a stochastic dichotomic variable $\eta_{jk}$, which can assume randomly
the values $\eta_{jk} = 0$ or $\eta_{jk} = 1$.
The model is defined by Eqs.~(\ref{basic0}) with
\begin{eqnarray}
  \Delta x =  - \eta_{jk} \omega x_k + (1-\eta_{jk}) \omega x_j \, .
  \label{basic0-A2}
\end{eqnarray}
The model describes a unidirectional flow of wealth from agent $k$ toward
agent $j$ for $\eta_{jk} = 1$ or vice versa for  $\eta_{jk} = 0$.
For the particular case in which the two values of $\eta_{jk}$ are 
always equiprobable, one can rewrite the process, without loss of generality,
with a $\Delta x = \omega x_k$ in Eqs.~(\ref{basic0}).
Numerical simulations of this model confirm the findings of Refs.~\cite{Angle2002a,Angle2006a},
that for small enough $\omega$ the stationary wealth distribution is well fitted
by a  $\mathrm{\Gamma}$-di\-stri\-bu\-tion $\gamma_n(x)$, 
with $n \approx 1/\omega - 1 = \lambda/(1 - \lambda)$.
We find that this fitting (not shown) is very good at least up to $\lambda \approx 0.7$.

\subsection{B-model}  \label{bennati}

\begin{figure}
\centering
\resizebox{0.75\columnwidth}{!}{
\includegraphics{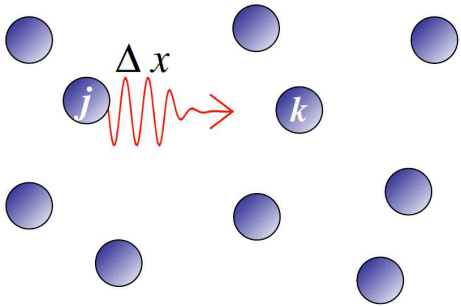}
}
\caption{
In the B-model the quantity $x$ can only vary by 
a constant amount $\Delta x = \Delta x_0$, 
which can e.g. be lost by a unit $j$ and then absorbed by a unit $k$, 
analogously to the emission-absorption process 
of light quanta of constant frequency.
}
\label{bennatimodel}
\end{figure}

Another KWEM was introduced in 1988 
by Eleonora Bennati~\cite{Bennati1988a,Bennati1988b}.
Its basic version, that we discuss here, 
is a simple unidirectional model where units exchange constant amounts 
$\Delta x_0$ of wealth~\cite{Bennati1988a,Bennati1988b,Bennati1993a}. 
In principle, in the B-model a situation where the wealths of the agents 
would become negative could occur.
This is prevented allowing the transaction to take place only 
if the condition $x_j', x_k' \ge 0$ is fulfilled,
i.e., the process is described by Eq.~(\ref{basic0}) 
with $\Delta x = \Delta x_0$
if $x_j', x_k' \ge 0$ and with  $\Delta x = 0$ otherwise.
Since the wealth can vary only by a constant amount $\Delta x_0$, 
the model reminds a set of particles exchanging energy 
by emitting and re-absorbing light quanta,
as illustrated symbolically in Fig.~\ref{bennatimodel}.
Analytically the equilibrium state of the B-model is well described 
by the exponential distribution (\ref{exp}).
A main difference respect to the other models considered here is that 
in the  B-model the amount of wealth exchanged between the two agents 
is independent of $x_i$, 
while in the other models represents a multiplicative random process,
since $\Delta x \propto x_i$.

\subsection{C-model} \label{CandC}

In the model introduced in 2000 by A. Chak\-ra\-bor\-ti 
and B. Chak\-ra\-bar\-ti~\cite{Chakraborti2000a} 
the general exchange rule reads,
\begin{eqnarray}
  x_j' &=& \lambda x_j + \epsilon \, (1-\lambda) (x_j + x_k) \, ,
  \nonumber \\
  x_k' &=& \lambda x_k + \bar{\epsilon} \, (1-\lambda) (x_j + x_k) \, ,
  \label{basic_CC}
\end{eqnarray}
where $\bar\epsilon = 1 - \epsilon$.
Here the new wealth $x_j'$ ($x_k'$) is expressed as a sum of
the saved fraction $\lambda x_j'$ ($\lambda x_k'$) of the initial wealth 
and a random fraction $\epsilon$ ($\bar{\epsilon}$) 
of the total remaining wealth, 
obtained summing the respective contributions of agents $j$ and $k$. 
Equations~(\ref{basic_CC}) are equivalent to Eqs.~(\ref{basic0}), with
\begin{eqnarray} \label{dx_CC}
\Delta x = \omega (\bar{\epsilon} \, x_j - \epsilon \, x_k )
=   (1 - \lambda) (\bar{\epsilon} \, x_j - \epsilon \, x_k)  \, .
\end{eqnarray}
Like in the A1-model, at equilibrium the system is well described by 
a $\mathrm{\Gamma}$-distribution (\ref{JA1}). 
For the parameter $n$ we find now~\cite{Patriarca2004a,Patriarca2004b}
\begin{eqnarray} \label{CC2}
n \equiv \frac{D}{2}
= \frac{1 + 2\lambda}{1 - \lambda} \,
= \frac{3}{\omega}  - 2 \, ,
\end{eqnarray}
which is twice the value of the corresponding parameter of the A1-model 
with $p_0 = 1/2$, discussed in Sec.~\ref{angle1}.

\begin{figure}
\resizebox{0.90\columnwidth}{!}{
\includegraphics{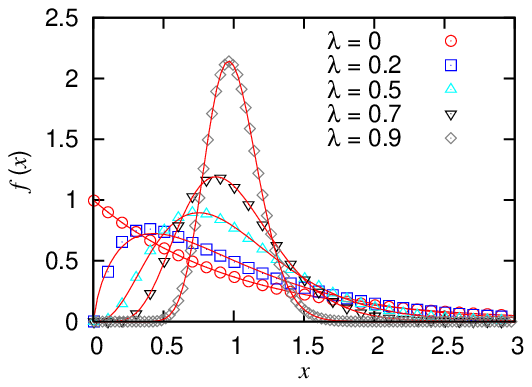}}\\
\resizebox{0.90\columnwidth}{!}{
\includegraphics{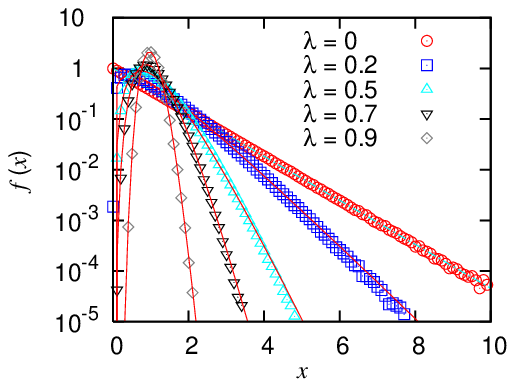}
}
\caption{
Equilibrium wealth distributions in linear and semi-log scale 
for different values of the saving parameter $\lambda$
in the closed economy model defined by Eqs.~(\ref{basic_CC}).
The continuous curves are the fitting functions, 
i.e. a $\mathrm{\Gamma}$-distribution
of the form Eq.~(\ref{JA1}) just as for the A1-model,
but the value of $n(\lambda)$ is given by Eq.~(\ref{CC2}).
}
\label{fig2}
\end{figure}

In Fig.~\ref{fig2} numerical results are compared with 
the fitting based on Eq.~(\ref{CC2}).
In this case the probability density is always finite for $x \to 0$, 
since for $\lambda = 0$ ($\omega = 1$) 
one has $n = 1$ and the distribution does not diverge,
being equal to the exponential function (\ref{exp}).
%

\subsection{D-model}
\label{dragu}

The models introduced in 2000 by A. Dra\-gu\-le\-scu 
and V. M. Ya\-ko\-ven\-ko~\cite{Dragulescu2000a} were conceived 
to describe flow and distribution of money.
They have a sound interpretation both of the conservation law 
$x_j' + x_k' = x_j + x_k$, since money is measured in the same unit 
and conserved during transactions, and of the stochasticity of the update rule,
representing a randomly chosen realization of trade.
Various models were considered in Ref.~\cite{Dragulescu2000a},
with a $\Delta x$ either constant (similarly to the B-model discussed above)
or dependent on the values $x_i$ of the agents;
also more realistic models, in which e.g. firms were introduced
or debts were allowed.
For simplicity, we consider among them the model which probably best represents
the random character of KWEMs, referred to as the D-model below,
in which the total initial amount $x_j + x_k$
is reshuffled randomly between the two interacting units,
\begin{eqnarray}
  x_j' &=& \epsilon \, (x_j + x_k) \, ,
  \nonumber \\
  x_k' &=& \bar{\epsilon} \, (x_j + x_k) \, .
  \label{basic1}
\end{eqnarray}
Equivalently, the dynamics can be described by Eqs.~(\ref{basic0}), with
\begin{eqnarray}
\Delta x = \bar{\epsilon} \, x_j - \epsilon \, x_k \, .
\end{eqnarray}
The D-model is formally recovered from the C-model 
for $\lambda = 0$ ($\omega = 1$).

\begin{figure}[t]
\centering
\resizebox{0.7\columnwidth}{!}{
\includegraphics{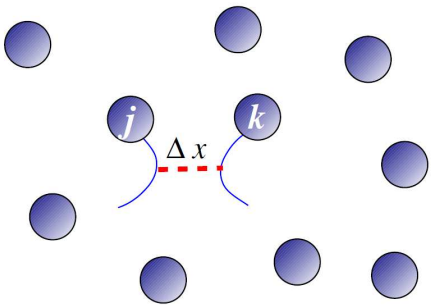}
}
\caption{
The D-model as well as the C-model prescribe 
a microscopic interaction between two units
analogously to a kinetic model of gas in which, 
during an elastic collision, two generic particles $j$ and $k$ 
exchange an energy amount $\Delta x$.
}
\label{CC}
\end{figure}

The equilibrium distribution of the D-model is well fitted by
the exponential distribution (\ref{exp}).
A mechanical analogue of the D-model is a gas, 
in which particles undergo pair collisions in which
some energy is exchanged~\cite{Whitney1990a},  
as symbolically illustrated in Fig.~\ref{CC}.
%

\subsection{Stationary wealth distributions}
\label{gamma}

The parameters of the $\mathrm{\Gamma}$-distribution, 
obtained from the fitting of the wealth distributions
of the stationary solutions for the models considered, 
are summarized in Table~\ref{table1}.
The analytical forms of the respective parameters $n$,
given as a function of  $\omega$ or $\lambda$, 
provide a good fitting: 
for the model A2, the fitting is good only up to $\lambda \approx 0.7$.
The close analogies among the various models are evident, 
however the existence of a general solution has not been demonstrated,
see e.g. Refs.~\cite{Repetowicz2005a,Angle2006a}.

\begin{table}[th]
\centering
\caption{Comparison of dependence of the fitting parameter $n$ in the $\mathrm{\Gamma}$-distribution 
(\ref{JA1}) on $\lambda$ or $\omega$, for the basic homogeneous versions 
of the A-, C-, and D-models.}
\label{table1}
\begin{tabular}{lccc}
\hline\noalign{\smallskip}
Model & $n(\omega)$ &  $n(\lambda)$ \\
\noalign{\smallskip}\hline\noalign{\smallskip}
A1    & $3/2\omega\!-\!1$ 
      & $(1\!+\!2\lambda)/2(1\!-\!\lambda)$ \\
A2    & $1/\omega\!-\!1$ 
      & $\lambda/(1\!-\!\lambda)$ \\
C     & $3/\omega\!-\!2$
      & $(1\!+\!2\lambda)/(1\!-\!\lambda)$ \\
D     & $1$
      & $1$ \\
\noalign{\smallskip}
\hline
\end{tabular}
\end{table}

\section{Influence of heterogeneity}
\label{heterogeneous}

Here we discuss the influence of heterogeneity,
considering as an example the generalization of the C-model.
Heterogeneity is introduced by assigning a different parameter 
$\omega_i$ ($\lambda_i$) to each agent $i$.
The formulation of the heterogeneous models 
can be straight\-for\-war\-dly obtained from those
of the corresponding homogeneous ones by replacing 
the generic term $\omega \, x_i$ ($\lambda x_i$) 
with $\omega_i x_i$ ($\lambda_i x_i$) in the evolution law.
In the case of the C-model Eqs.~(\ref{basic_CC}) become
\begin{eqnarray}
  x_j' 
  &=& 
  \lambda_j x_j 
  + \epsilon \,  [(1-\lambda_j) x_j + (1-\lambda_k) x_k] \, ,
  \nonumber \\
  x_k' 
  &=& 
  \lambda_k x_k 
  + \bar{\epsilon} \, [(1-\lambda_j) x_j + (1-\lambda_k) x_k] \, ,
  \label{basic_CC1}
\end{eqnarray}
and the exchanged amount of wealth in Eqs.~(\ref{basic0}) is now
\begin{eqnarray} \label{dx_CC1}
\Delta x 
=  \bar{\epsilon} \, \omega_j \, x_j \!-\! \epsilon \, \omega_k \, x_k
=  \bar{\epsilon} (1 \!-\! \lambda_j) x_j \!-\! \epsilon (1 \!-\! \lambda_k) x_k
\, .~
\end{eqnarray}
The set of parameters $\{\omega_i\}$ ($\{\lambda_i\}$) 
is constant in time and specifies the profiles of the agents.
The values $\{\omega_i\}$ ($\{\lambda_i\}$) are assumed to be distributed 
in the interval between $0$ and $1$ with probabilities $ h_i$ ($g_i$) and 
$\sum_i h_i\!=\!1$ ($\sum_i g_i\!=\!1$).
In the limit of an infinite number of agents, one can introduce
a probability distribution $ h(\omega)$ [$g(\lambda)$],
with $\int_0^1 \! d\omega \,  h(\omega) \!=\! 1$
[$\int_0^1 \! d\lambda \, g(\lambda) \!=\! 1$].

Various analytical and numerical studies 
of this model have been carried out~\cite{Chatterjee2005b,Chatterjee2007b,Angle2002a,Iglesias2004a,Repetowicz2005a,Chatterjee2003a,Das2003a,Chatterjee2004a,Chatterjee2005a,Das2005a,Patriarca2005a-brief,Silva2005a} 
and as a main result it has been found that 
the exponential law remains limited to intermediate $x$-values, 
while a Pareto power law appears at larger values  of $x$.
Such a shape is prototypical for real wealth distributions.
Numerical simulations and theoretical considerations suggest 
that the power law exponent is quite insensitive to the details
of the system parameters, i.e., to the distribution $ h(\omega)$.
In fact, the Pareto exponent depends 
on the limit $g(\lambda\!\to\!1)$.
If $g(\lambda) \sim (1-\lambda)^{\alpha-1}$ with $\lambda \to 1$
and $\alpha \le 1$, then the corresponding power law has 
an exponent $\alpha$~\cite{Chatterjee2004a}.
Thus, in general, agents with $\lambda_i$ close to $1$ 
are responsible for the appearance of the power law tail~\cite{Chatterjee2004a,Patriarca2005a-brief,Patriarca2006c}.

Probably the most interesting feature of the equilibrium state
is that while the shape of the wealth distribution $f_i(x)$ 
of agent $i$ is a $\mathrm{\Gamma}$-distribution, 
the sum of the wealth distributions of the single agents,
$f(x) = \sum_i f_i(x)$, produces a power law tail.
Vice versa, one could say that the global wealth distribution $f(x)$ 
can be resolved as a mixture of partial wealth probability densities $f_i(x)$ 
with exponential tail, with different parameters.
For instance, the corresponding average wealth depends on the saving parameter 
as $\langle x \rangle_i \propto 1/(1-\lambda_i) = 1/\omega_i$;
see Refs.~\cite{Bhattacharya2005a-brief,Patriarca2005a-brief,Patriarca2006c}
for details.

Importantly, all real distributions have a finite cutoff;
no real wealth distribution has an infinitely extended power law tail.
The Pareto law is always observed between a minimum wealth value $x_\mathrm{min}$
and a cutoff $x_\mathrm{max}$, representing the wealth of the richest agent.
This can be well reproduced by the heterogeneous model using 
an upper cutoff $\lambda_{\rm max} < 1$ for the saving parameter distribution 
$g(\lambda)$: the closer to one is $\lambda_{\rm max}$,
the larger is $x_\mathrm{max}$ and wider the interval in which 
the power law is observed~\cite{Patriarca2006c}.

The role of the $\lambda$-cutoff is closely related to and
relevant for understanding the relaxation process.
The relaxation time scales of single agents in a heterogeneous model
are proportional to $1/(1 - \lambda_i)$~\cite{Patriarca2007a}.
This means that the slowest convergence rate is determined 
by $1 - \lambda_\mathrm{max}$.
In numerical simulations of heterogeneous KWEMs,
one necessarily employes a finite $\lambda$-cutoff.
However, this should not be regarded as a limit of numerical simulations but
a feature suited to describe real wealth distributions.
Simulations confirm the fast convergence to equilibrium for each agent 
with the above mentioned time scale~\cite{Patriarca2007a}.
Gupta has demonstrated numerically that the convergence 
is exponentially fast~\cite{Gupta2008a}.

In Ref.~\cite{During2008a} it has been claimed that heterogeneous KWEMs 
with randomly distributed $\lambda_i$ ($0 \le \lambda_i < 1$)
cannot undergo a fast relaxation toward an equilibrium wealth distribution,
but the relaxation should instead take place on algebraic time scale.
This in turn means that there cannot exists any power law tail.
Such claims are probably correct for systems with a $\lambda$-distribution
$g(\lambda)$ rigorously extending as far as $\lambda=1$,
corresponding to a power law tail extending as far as $x = \infty$.
However, this does not apply to KWEMs with a saving parameter cutoff
$\lambda_\mathrm{max} < 1$, which is the natural choice
in describing real systems, as well as in numerical simulations,
employing a finite $\lambda$-cutoff:
in this case the largest time scale is finite and relaxation is fast.

\section{Generalizations}
\label{reformulation}

In this section a unified reformulation of the exchange laws 
of the A-, C-, and D-models is suggested and 
as an example an application to the C-model is made.

\subsection{Reformulation}
\label{formula}

It is possible to reformulate the evolution law either
through a single stochastic saving variable $\tilde{\lambda}$ 
or an equivalent stochastic exchange variable 
$\tilde\omega = 1 - \tilde{\lambda}$.
This formal rearrangement of the equations maintains 
the form of the evolution law very simple and  
has at the same time the advantage to be particularly suitable 
to make further generalizations.
For the sake of generality, we consider the case of a heterogeneous system 
characterized by a parameter set $\{\omega_i\}$.
The models discussed above (apart from the B-model) 
can be rewritten according to the basic equations (\ref{basic0}),
where the wealth exchange term is now given by
\begin{eqnarray}
  \Delta x = \tilde\omega_j \, x_j - \tilde \omega_k \, x_k \, .
  \label{ref0}
\end{eqnarray}
The meaning of the new stochastic variables $\tilde\omega_j$
and $\tilde\omega_k$ introduced is simple: 
$\tilde\omega_j$ represents the fraction of wealth given by agent $j$ to $k$
during the transaction, and vice versa for $\tilde\omega_k$.
Comparison with the equations defining the A-, C-, and D-models provides
the following definitions for $\tilde\omega_j$ and $\tilde\omega_k$:

\begin{itemize}
\item
In the A1-model, $\tilde \omega_j$ and $\tilde \omega_k$ are 
independent nonlinear stochastic functions of the agent wealths $x_j$ and $x_k$,
\begin{eqnarray}
  \label{refA}
  \tilde \omega_j &=& \eta_{j,k} \epsilon \, \omega_j \, ,
  \nonumber \\	
  \tilde \omega_k &=& (1  - \eta_{j,k}) \epsilon \, \omega_k \, ,
\end{eqnarray}
where $\eta_{j,k} = \phi(x_k-x_j) = 1$ 
with probability $p_0$ for $x_k - x_j > 0$ 
and $\eta_{j,k} = 0$ with probability $1-p_0$ for $x_k - x_j < 0$, 
while $\epsilon$ is a random number in $(0,1)$.
For $\eta_{j,k} = 0$ one has $\tilde \omega_j = 0$ 
and $\tilde \omega_k \in (0, \omega_k)$, whereas for 
$\eta_{j,k} = 1$ one has $\tilde \omega_k = 0$ 
and $\tilde \omega_j \in (0, \omega_j)$.
\item
In the A2-model, $\tilde \omega_j$ and $\tilde \omega_k$
only contain the dichotomic variable,
\begin{eqnarray}
  \label{refA2}
  \tilde \omega_j &=& \eta_{j,k} \omega_j \, ,
  \nonumber \\	
  \tilde \omega_k &=& (1  - \eta_{j,k}) \omega_k \, .
\end{eqnarray}
%
%
%
\item
For the C-model,
\begin{eqnarray}
\label{refC1}
  \tilde \omega_j 
  &=&  \epsilon \, \omega_j \, ,
  ~~~~~~~~~~~~~~~\,\!\tilde \omega_j \in (0, \omega_j) \, ,
    \nonumber \\	
  \tilde \omega_k 
  &=& (1-\epsilon) \, \omega_k \, ,
  ~~~~~~~\tilde \omega_k \in (0, \omega_k) \, ,
\end{eqnarray}
where $\epsilon$ is a random number in $(0,1)$.
%
%
%
%
\item
The D-model is recovered from Eqs.~(\ref{refC1}) of the C-model 
when $\omega_i = 1$ for each agent $i$.
\end{itemize}

\begin{table}[th]
\centering
\caption{Comparison of the A-, C-, and D-models:
explicit forms of the random variables 
$\tilde\omega_j$ and $\tilde \omega_k$
in the unified reformulation~(\ref{ref0}).
See text for details.}
\label{table2}
\begin{tabular}{lcc}
\hline\noalign{\smallskip}
Model & $\tilde\omega_j$
      & $\tilde \omega_k$\\
\noalign{\smallskip}\hline\noalign{\smallskip}
A1    & $\epsilon\,\eta\,\omega_j$
      & $(1\!-\!\eta)\epsilon\,\omega_k$\\
A2    & $\eta\,\omega_j$
      & $(1\!-\!\eta) \omega_k$\\
C     & $\epsilon\,\omega_j$
      & $(1\!-\!\epsilon)\,\omega_k$\\
D     & $\epsilon$
      & $(1\!-\!\epsilon)$\\
\noalign{\smallskip}
\hline
\end{tabular}
\end{table}
The reformulation is summarized in Table \ref{table2}
with reference to Eq.~(\ref{ref0}).
Different generalizations can now be done changing only the properties 
of the stochastic variables $\tilde \omega_i$, 
while maintaining the same formulation (\ref{ref0}) of the exchange law.

\subsection{An example}
\label{example}

\begin{figure}
\resizebox{0.90\columnwidth}{!}{
\includegraphics{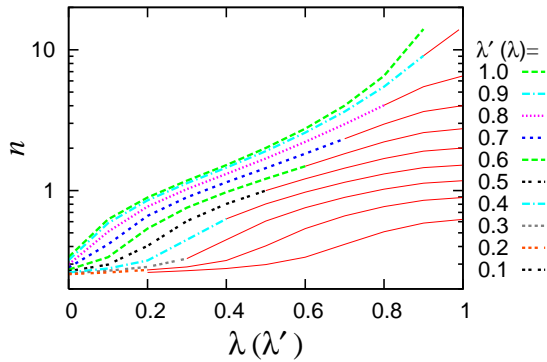}
}
\caption{
  Fitting parameter $n(\lambda,\lambda')$ 
  of the $\mathrm{\Gamma}$-distribution $\gamma_n(x)$ 
  for the generalized model with $\lambda < \tilde \lambda_i < \lambda'$.
  Dotted and dashed lines: $n$ as a function of $\lambda$
  ($\le \lambda'$ by definition)
  for fixed values of $\lambda'$ shown on the right side.
  These lines stop at $\lambda = \lambda'$ since by definition 
  $\lambda \le \lambda'$.
  Continuous lines: $n$ as a function of $\lambda'$ (($\ge \lambda$) 
  for the same fixed values of $\lambda$ shown on the right side.
}
\label{n-L}
\end{figure}

As an example which can be represented through 
Eq.~(\ref{ref0}), we consider a generalization of the homogeneous C-model.
In the original version there is a constraint on the maximum fraction 
of invested wealth, given by a value $0 < \omega \le 1$ of the exchange parameter,
or equivalently on the minimum saved fraction,
given by a value $0 \le \lambda < 1$ of the saving parameter.
Now an additional constraint on the minimum fraction of the invested wealth
is assumed.
This may describe e.g. trades which always have a minimum risk for an agent.  
It can be represented by an analogous parameter $\omega'$, 
with $0 < \omega' < \omega$, representing the minimum fraction of wealth 
invested in a single trade.
One can also define a parameter $\lambda' = 1 - \omega'$,
with $\lambda < \lambda' < 1$, representing the maximum fraction of saved
wealth (i.e. it is not possible to go through a trade
without risking a non-zero amount of wealth).
Then the stochastic variables $\tilde \omega_i$ in Eq.~(\ref{ref0}) 
become uniform random numbers in intervals defined
by the parameters $\omega'$ and $\omega$ (or by $\lambda'$ and $\lambda$),
\begin{eqnarray}
  \tilde \omega_i \in (\omega', \omega) = (1-\lambda', 1-\lambda) \, .
\end{eqnarray}
We have performed numerical simulations for a set of combinations of
parameters $(\lambda,\lambda')$ and found that the equilibrium distributions 
are always well fitted by the same $\mathrm{\Gamma}$-distribution (\ref{JA2}).
However, we have not found a simple analytical formula for fitting 
the dependence of the parameter $n$ on the saving parameters 
$\lambda $ and $\lambda'$.
The behavior of $n$ {\it versus} $\lambda$ ($\lambda'$) 
is represented graphically in Fig.~\ref{n-L}.
Dotted/dashed curves (different colors) represent $n$ {\it versus} $\lambda$ 
for the different fixed values of $\lambda'$ shown on the right side
These curves stop at $\lambda = \lambda'$,
since by definition $\lambda < \lambda'$.
From there the continuous (red) curves 
start, which represent $n$ {\it versus} $\lambda'$ for the same fixed values 
of $\lambda$  listed in the legend on the right.
The first (dashed green) curve from the top extending on the whole interval 
$\lambda = (0,1)$ represents $n$ as a function of $\lambda$ f
or $\lambda' = 1$ and corresponds to the original 
homogeneous C-model.
For this particular case $n(\lambda)$ is known to diverge 
as $n \sim 1/(1-\lambda) \sim 1/\omega$ for $\lambda \to 1$ 
(see Table~\ref{table1}),
while in all the other cases $n$ is finite.
%

\section{Conclusions and discussion}
\label{conclusion}

We have reviewed some basic KWEMs of closed economy systems, 
introduced by scientists working in different fields,
allowing us to point out analogies and differences between them.
We have first considered the homogeneous models and 
then discussed the influence of heterogeneity.
The heterogeneous KWEMs are particularly relevant in the study of real wealth 
distributions, since they can reproduce both the exponential shape at 
intermediate values of wealth as well as the power law tail.

In all the models discussed, including the heterogeneous one, 
the equilibrium wealth distribution of a single agent is well fitted 
by a $\mathrm{\Gamma}$-distribution, known to be the canonical distribution 
of a general harmonic system with a suitable number of degrees of freedom.
This suggests a simple mechanism underlying the
(approach to) equilibrium of these systems,
similar to the energy redistribution in a mechanical system.
However, a general demonstration that the $\mathrm{\Gamma}$-distribution
is the stationary solution of KWEMs and an understanding of how it arises is still missing
(see Refs.~\cite{Repetowicz2005a,Angle2006a,ChakraPat2009,Anindya2009} 
for theoretical considerations on and the microeconomic formulation of this issue).

Furthermore, we have discussed how in a heterogeneous KWEM the sum of the single agent
wealth distributions can produce a power law tail.
In particular, we have clarified some issues concerning the relaxation process 
and the existence of power law tails:
whenever there is a finite cutoff in the saving parameter distribution,
the largest time scale of the system is finite and one observes
a fast (exponential) relaxation toward a power law, which extends over
a finite interval of wealth.
The width of such interval depends on saving parameter cutoff.

Due to the similarity of the structures of the models discussed, 
we have proposed a novel unified reformulation based on the introduction of
suitable stochastic variables $\tilde \omega_i$, 
representing the \emph{actual fraction of wealth lost
by the $i$-th agent} during a single transaction.
This unified formulation lends itself easily to further generalizations,
which can be obtained by modifying the stochastic properties 
of the variable $\tilde \omega_i$ only, 
while leaving the general evolution law unchanged.
We have illustrated the new formulation by working out in detail an example,
in which the fraction of wealth lost $\tilde \omega_i$ 
is characterized by a lower as well as an upper limit.

Besides the KWEMs considered in the present paper, 
originally formulated through finite time difference stochastic equations,
other relevant (versions of) KWEMs have been introduced in the literature; 
see Refs.~\cite{Lux2005a,Chatterjee2007b,Yakovenko2009a} for an overview.
Their mathematical formulation can be similar to the one of the present 
paper~\cite{Angle1986a,Angle2002a,Angle2006a,Iglesias2004a,Ausloos2007a}, 
or different approaches can be used, such as matrix theory~\cite{Gupta2006a},
the master equation~\cite{Ispolatov1998a,Bouchaud2000b,Ferrero2004a},
the Boltzmann equation~\cite{Slanina2004a,Repetowicz2005a,Cordier2005a,Matthes2007a,During2007a,During2008a},
the Lotka-Volterra equation~\cite{Solomon2001a,Solomon2002a},
or Markov chains models~\cite{Scalas2006b,Scalas2007a,Garibaldi2007a}.
All these models share a description of wealth flow as due to exchanges between
basic units.
In this respect, they are all very different from the class of models
formulated in terms of a Langevin equation for a single wealth variable 
subjected to multiplicative noise~\cite{Gibrat1931a,Mandelbrot1960a,Levy1996a,Sornette1998a,Burda2003a}.
The latter models can lead to wealth distributions with a power law tail.
In fact, they converge toward a log-normal distribution,
which, however, does not fit real wealth distributions 
as well as a $\mathrm{\Gamma}$-distribution or a $\mathrm{\beta}$-distribution
and is asymptotically characterized by too large variances~\cite{Angle1986a}.

Finally, we would like to point out that even though 
KWEMs have been the subject of intensive investigations,
their economical interpretation is still an open problem.
It is important to keep in mind that in the framework of a KWEM 
the agents should not be related to the {\it rational agents} 
of neoclassical economics:
an interaction between two agents does not represent
the effect of decisions taken by two economic agents who 
have full information about the market
and behave rationally in order to maximize their utility.
The description of wealth flow provided by KWEMs
takes into account the stochastic element,
which does not respond by definition to any rational criterion.
Also some terms employed in the study of KWEMs, 
such as {\it saving propensity} (replaced here by {\it saving parameter}), 
{\it risk aversion}, etc., can be misleading since they seem to imply 
a decisional aspect behind the behavior of agents.
Trying to interpret the dynamics of KWEMs through concepts taken 
from the neoclassical theory leads to obvious misunderstandings~\cite{Lux2008a}.
However, it is interesting to note that very recently, Chakrabarti and Chakrabarti 
have put forward a microeconomic
formulation of the above models, using the utility function as a guide to the
behavior of agents in the economy~\cite{Anindya2009}.
Instead, KWEMs provide a description at a coarse grained level,
as in the case of many statistical mechanical models,
where the connection with the microscopic mechanisms
is not visible; however, the equivalence is maintained.

\section*{Acknowledgments}

This work has been supported 
by the EU NoE BioSim, LSHB-CT-2004-005137 (M.P.), 
Spa\-nish MI\-CINN and FE\-DER 
through project FISICOS (FIS2007-60327) (E.H.), 
Estonian Ministry of Education and Research
through Project No. SF0690030s09, 
and Estonian Science Foundation via grant no. 7466 (M.P., E.H.).



\begin{thebibliography}{10}

\expandafter\ifx\csname url\endcsname\relax
  \def\url#1{\texttt{#1}}\fi
\expandafter\ifx\csname urlprefix\endcsname\relax\def\urlprefix{URL }\fi

\bibitem{Bouchaud2005a}
J.-P. Bouchaud, The subtle nature of financial random walks, CHAOS 15 (2005)
  026104.

\bibitem{Chatterjee2005b}
A.~Chatterjee, S.~Yarlagadda, B.~K. Chakrabarti (Eds.), Econophysics of Wealth
  Distributions - Econophys-Kolkata I, Springer, 2005.

\bibitem{Feigenbaum2003a}
J.~Feigenbaum, Financial physics, Rep. Prog. Phys. 66 (2003) 1611.

\bibitem{Mantegna2000a}
R.~N. Mantegna, H.~E. Stanley, An Introduction to Econophysics, Cambridge
  University Press, Cambridge, 2000.

\bibitem{Stauffer2005a}
D.~Stauffer, C.~Schulze, Microscopic and macroscopic simulation of competition
  between languages, Phys. Life Rev. 2 (2005) 89.

\bibitem{Stauffer2006c}
D.~Stauffer, S.~M. de~Oliveira, P.~M.~C. de~Oliveira, J.~S. de~Sa~Martins,
  Biology, Sociology, Geology by Computational Physicists, Elsevier Science,
  2006.

\bibitem{Chakrabarti2006a}
B.~K. Chakrabarti, A.~Chakraborti, A.~Chatterjee (Eds.), Econophysics and
  Sociophysics: Trends and Perspectives, 1st Edition, Wiley - VCH, Berlin,
  2006.

\bibitem{Taagepera2008b}
R.~Taagepera, Making social sciences more scientific. The need for predictive
  models, Oxford University Press, Oxford, 2008.

\bibitem{Ball2002a}
P.~Ball, The physical modelling of society: a historical perspective, Physica A
  314 (2002) 1.

\bibitem{Yakovenko2009a}
V.~Yakovenko, J.~J.~Barkley~Rosser, Statistical mechanics of money, wealth, and
  income, arXiv:0905.1518.
\newline\urlprefix\url{www.arxiv.org}

\bibitem{Pareto1897a}
V.~Pareto, Cours d'economie politique, Rouge, Lausanne, 1897.

\bibitem{Pareto1971a_reprint}
V.~Pareto, Manual of political economy, Kellag, New York, 1971.

\bibitem{Bachelier1900a}
L.~Bachelier, Theorie de la speculation, Annales Scientifiques de l'Ecole
  Normale Superieure III-17 (1900) 21.

\bibitem{Bachelier1990b}
{E}nglish translation of: {L}. {B}achelier, {T}heorie de la speculation,
  {A}nnales {S}cientifiques de l'\'{E}cole {N}ormale {S}uperieure (1900) {III}
  -17, in: S.~Haberman, T.~A. Sibbett (Eds.), History of Actuarial Science,
  Vol.~7, Pickering and Chatto Publishers, London, 1995, p.~15.

\bibitem{Boness1967a}
A.~J. Boness, {E}nglish translation of: {L}. {B}achelier, {T}heorie de la
  {S}peculation, {A}nnales de l'{E}cole {N}ormale {S}uperieure {III}-17 (1900),
  pp. 21-86, in: P.~H. Cootner (Ed.), The Random Character of Stock Market
  Prices, MIT, Cambridge, MA, 1967, p.~17.

\bibitem{Mandelbrot1963a}
B.~B. Mandelbrot, The variation of certain speculative prices, J. Business 36
  (1963) 394.

\bibitem{Arthur1999a}
W.~B. Arthur, Science 284 (1999) 107.

\bibitem{Hayes2002a}
B.~Hayes, Follow the money, Am. Sci. 90~(5).

\bibitem{Chatterjee2007b}
A.~Chatterjee, B.~Chakrabarti, Kinetic exchange models for income and wealth
  distributions, Eur. Phys. J. B 60 (2007) 135.

\bibitem{Angle1983a}
J.~Angle, The surplus theory of social stratification and the size distribution
  of personal wealth, in: Proceedings of the {A}merican {S}ocial {S}tatistical
  {A}ssociation, {S}ocial {S}tatistics {S}ection, Alexandria, VA, 1983, p. 395.

\bibitem{Angle1986a}
J.~Angle, The surplus theory of social stratification and the size distribution
  of personal wealth, Social Forces 65 (1986) 293.
\newline\urlprefix\url{http://www.jstor.org}

\bibitem{Angle2002a}
J.~Angle, The statistical signature of pervasive competition on wage and salary
  incomes, J. Math. Sociol. 26 (2002) 217.

\bibitem{Angle2006a}
J.~Angle, The inequality process as a wealth maximizing process, Physica A 367
  (2006) 388.

\bibitem{Bennati1988a}
E.~Bennati, La simulazione statistica nell'analisi della distribuzione del
  reddito: modelli realistici e metodo di {M}onte {C}arlo, ETS Editrice, Pisa,
  1988.

\bibitem{Bennati1988b}
E.~Bennati, Un metodo di simulazione statistica nell'analisi della
  distribuzione del reddito, Rivista Internazionale di Scienze Economiche e
  Commerciali 35 (1988) 735.

\bibitem{Bennati1993a}
E.~Bennati, Il metodo {M}onte {C}arlo nell'analisi economica, Rassegna di
  lavori dell'ISCO X (1993) 31.

\bibitem{Chakraborti2000a}
A.~Chakraborti, B.~K. Chakrabarti, Statistical mechanics of money: {H}ow saving
  propensity affects its distribution, Eur. Phys. J. B 17 (2000) 167.

\bibitem{Dragulescu2000a}
A.~Dragulescu, V.~M. Yakovenko, Statistical mechanics of money, Eur. Phys. J. B
  17 (2000) 723.

\bibitem{Mandelbrot1960a}
B.~Mandelbrot, The {P}areto-{L}evy law and the distribution of income, Int.
  Econ. Rev. 1 (1960) 79.

\bibitem{Levy1997a}
M.~Levy, S.~Solomon, New evidence for the power-law distribution of wealth,
  Physica A 242 (1997) 90.

\bibitem{Fujiwara2003a}
Y.~Fujiwara, W.~Souma, H.~Aoyama, T.~Kaizoji, M.~Aoki, Growth and fluctuations
  of personal income, Physica A 321 (2003) 598.

\bibitem{Aoyama2003a}
H.~Aoyama, W.~Souma, Y.~Fujiwara, Growth and fluctuations of personal and
  company's income, Physica A 324 (2003) 352.

\bibitem{Dragulescu2001a}
A.~Dragulescu, V.~M. Yakovenko, Exponential and power-law probability
  distributions of wealth and income in the {U}nited {K}ingdom and the {U}nited
  {S}tates, Physica A 299 (2001) 213.

\bibitem{Dragulescu2001b}
A.~Dragulescu, V.~M. Yakovenko, Evidence for the exponential distribution of
  income in the usa, Eur. Phys. J. B 20 (2001) 585.

\bibitem{Ferrero2004a}
J.~C. Ferrero, The statistical distribution of money and the rate of money
  transference, Physica A 341 (2004) 575.

\bibitem{Iglesias2004a}
J.~R. Iglesias, S.~Goncalves, G.~Abramsonb, J.~L. Vega, Correlation between
  risk aversion and wealth distribution, Physica A 342 (2004) 186.

\bibitem{Patriarca2007a}
M.~Patriarca, A.~Chakraborti, E.~Heinsalu, G.~Germano, Relaxation in
  statistical many-agent economy models, Eur. J. Phys. B 57 (2007) 219.

\bibitem{Patriarca2004a}
M.~Patriarca, A.~Chakraborti, K.~Kaski, Statistical model with a standard gamma
  distribution, Phys. Rev. E 70 (2004) 016104.

\bibitem{Patriarca2004b}
M.~Patriarca, A.~Chakraborti, K.~Kaski, {G}ibbs versus non-{G}ibbs
  distributions in money dynamics, Physica A 340 (2004) 334.

\bibitem{Whitney1990a}
C.~A. Whitney, Random processes in physical systems. An introduction to
  probability-based computer simulations, Wiley Interscience, NY, 1990.

\bibitem{Repetowicz2005a}
P.~Repetowicz, S.~Hutzler, P.~Richmond, Dynamics of money and income
  distributions, Physica A 356 (2005) 641.

\bibitem{Chatterjee2003a}
A.~Chatterjee, B.~K. Chakrabarti, S.~S. Manna, Money in gas-like markets:
  {G}ibbs and {P}areto laws, Physica Scripta T 106 (2003) 367.

\bibitem{Das2003a}
A.~Das, S.~Yarlagadda, A distribution function analysis of wealth distribution.
\newline\urlprefix\url{arxiv.org:cond-mat/0310343}

\bibitem{Chatterjee2004a}
A.~Chatterjee, B.~K. Chakrabarti, S.~S. Manna, {P}areto law in a kinetic model
  of market with random saving propensity, Physica A 335 (2004) 155.

\bibitem{Chatterjee2005a}
A.~Chatterjee, B.~K. Chakrabarti, R.~B. Stinchcombe, Master equation for a
  kinetic model of trading market and its analytic solution, Phys. Rev. E 72
  (2005) 026126.

\bibitem{Das2005a}
A.~Das, S.~Yarlagadda, An analytic treatment of the {G}ibbs--{P}areto behavior
  in wealth distribution, Physica A 353 (2005) 529.

\bibitem{Patriarca2005a-brief}
M.~Patriarca, A.~Chakraborti, K.~Kaski, G.~Germano, Kinetic theory models for
  the distribution of wealth: {P}ower law from overlap of exponentials, in
  Ref.~\cite{Chatterjee2005b}  p.93.

\bibitem{Silva2005a}
A.~C. Silva, V.~M. Yakovenko, Temporal evolution of the `thermal' and
  `superthermal' income classes in the usa during 1983-2001, Europhysics
  Letters 69 (2005) 304.

\bibitem{Patriarca2006c}
M.~Patriarca, A.~Chakraborti, G.~Germano, Influence of saving propensity on the
  power law tail of wealth distribution, Physica A 369 (2006) 723.

\bibitem{Bhattacharya2005a-brief}
K.~Bhattacharya, G.~Mukherjee, S.~S. Manna, Detailed simulation results for
  some wealth distribution models in econophysics, in
  Ref.~\cite{Chatterjee2005b}  p.111.

\bibitem{Gupta2008a}
A.~K. Gupta, Relaxation in the wealth exchange models, Physica A 387 (2008)
  6819.

\bibitem{During2008a}
B.~D\"uring, D.~Matthes, G.~Toscani, Kinetic equations modelling wealth
  redistribution: A comparison of approaches, Phys. Rev. E 78 (2008) 056103.

\bibitem{ChakraPat2009}
A. Chakraborti and M. Patriarca, A variational principle for the Pareto law, 
arXiv:cond-mat/0605325v2 (2008).
\newline\urlprefix\url{www.arxiv.org}

\bibitem{Anindya2009}
A-S. Chakrabarti, B. K. Chakrabarti, Physica A, 388 (2009) 4151.

\bibitem{Lux2005a}
T.~Lux, Emergent statistical wealth distributions in simple monetary exchange
  models: A critical review, in: A.~Chatterjee, S.Yarlagadda, B.~K. Chakrabarti
  (Eds.), Econophysics of Wealth Distributions, Springer, 2005, p.~51.

\bibitem{Ausloos2007a}
M.~Ausloos, A.~Pekalski, Model of wealth and goods dynamics in a closed market,
  Physica A 373 (2007) 560.

\bibitem{Gupta2006a}
A.~K. Gupta, Money exchange model and a general outlook, Physica A 359 (2006)
  634.

\bibitem{Ispolatov1998a}
S.~Ispolatov, P.~L. Krapivsky, S.~Redner, Wealth distributions in asset
  exchange models, Eur. Phys. J. B 2 (1998) 267.

\bibitem{Bouchaud2000b}
J.~P. Bouchaud, M.~Mezard, Wealth condensation in a simple model of economy,
  Physica A 282 (2000) 536.

\bibitem{Slanina2004a}
F.~Slanina, Inelastically scattering particles and wealth distribution in an
  open economy, Phys. Rev. E 69 (2004) 046102.

\bibitem{Cordier2005a}
S.~Cordier, L.~Pareschi, G.~Toscani, On a kinetic model for a simple market
  economy, J. Stat. Phys. 120 (2005) 253.

\bibitem{Matthes2007a}
D.~Matthes, G.~Toscani, On steady distributions of kinetic models of
  conservative economies, J. Stat. Phys. 130 (2007) 1087.

\bibitem{During2007a}
B.~D\"uring, G.~Toscani, Hydrodynamics from kinetic models of conservative
  economies, Physica A 384 (2007) 493.

\bibitem{Solomon2001a}
S.~Solomon, P.~Richmond, Power laws of wealth, market order volumes and market
  returns, Physica A 299 (2001) 188.

\bibitem{Solomon2002a}
S.~Solomon, P.~Richmond, Stable power laws in variable economies;
  {L}otka-{V}olterra implies {P}areto-{Z}ipf, Eur. Phys. J. B 27 (2002) 257.

\bibitem{Scalas2006b}
E.~Scalas, U.~Garibaldi, S.~Donadio, Statistical equilibrium in simple exchange
  games {I}, Eur. Phys. J. B 53 (2006) 267.

\bibitem{Scalas2007a}
E.~Scalas, U.~Garibaldi, S.~Donadio, {\it Erratum}. Statistical equilibrium in
  simple exchange games {I}, Eur. Phys. J. B 60 (2007) 271.

\bibitem{Garibaldi2007a}
U.~Garibaldi, E.~Scalas, P.~Viarengo, Statistical equilibrium in simple
  exchange games {II}. the redistribution game, Eur. Phys. J. B 60 (2007) 241.

\bibitem{Gibrat1931a}
R.~Gibrat, Les In\'egalit\'es Economiques, Sirey, 1931.

\bibitem{Levy1996a}
M.~Levy, S.~Solomon, Power laws are logarithmic {B}oltzmann laws, Int. J. Mod.
  Phys. C 7 (1996) 595.

\bibitem{Sornette1998a}
D.~Sornette, Multiplicative processes and power laws, Phys. Rev. E 57 (1998)
  4811.

\bibitem{Burda2003a}
Z.~Burda, J.~Jurkiewics, M.~A. Nowak, Is {E}conophysics a solid science?, Acta
  Physica Polonica B 34 (2003) 87.

\bibitem{Lux2008a}
T.~Lux, Applications of statistical physics in finance and economics, Kiel
  Working Paper 1425.


\end{thebibliography}
\end{document}